\documentclass[aps,prl,twocolumn,floatfix]{revtex4-1}
\usepackage{natbib}
\usepackage{graphicx}
\usepackage{amssymb}
\usepackage{amsmath}
\usepackage{color}
\usepackage{multirow}
\begin{document}

\title{Nonlinear and magic ponderomotive spectroscopy}
\author{K.~R.~Moore*}
\author{G.~Raithel}
\affiliation{Applied Physics Program and Department of Physics, University of Michigan, Ann Arbor, MI 48109}
\date{\today }
\begin{abstract}
In ponderomotive spectroscopy an amplitude-modulated optical standing wave is employed to probe Rydberg-atom transitions, utilizing a ponderomotive rather than a dipole-field interaction.  Here, we engage nonlinearities in the modulation to drive dipole-forbidden transitions up to the fifth order.  We reach transition frequencies approaching the sub-THz regime.  We also demonstrate magic-wavelength conditions, which result in symmetric spectral lines with a Fourier-limited feature at the line center.  Applicability to precision measurement is discussed.
\end{abstract}
\maketitle

Measurements of atomic transition frequencies are the cornerstone of precision metrology, used in applications ranging from atomic clocks~\cite{Ye.2014} to measuring gravitational redshifts~\cite{Wineland.2010} and the radius of the proton~\cite{Pohl.2010}.  An important metric of precision in these applications is the fractional frequency resolution, $\Delta \nu / \nu$, in which $\nu$ is the measured frequency and $\Delta \nu$ is its uncertainty.  In order to obtain the best $\Delta \nu / \nu$, it is desirable to increase $\nu$ while decreasing $\Delta \nu$.  In recently-developed ponderomotive spectroscopy~\cite{Moore.2015}, Rydberg atoms are trapped in a standing-wave laser field (optical lattice).  Electronic transitions are driven by modulating the lattice-light intensity at the transition frequencies of interest.  In this Letter, we employ nonlinearities intrinsic to this excitation process to increase $\nu$ by driving atomic transitions at higher harmonics of the modulation frequency.  Unlike in standard nonlinear spectroscopy, in ponderomotive spectroscopy the intensity of the probe field does not need to be increased to drive these higher-order transitions, keeping light shifts low.  We also identify magic transitions that minimize residual trap-induced shifts, in addition to reducing the uncertainty $\Delta \nu$ of the line center by about one order of magnitude.

The ponderomotive interaction of a Rydberg electron with an optical field allows one to drive transitions when there is substantial spatial variation of the field intensity within the volume of the atom and when the lattice potential is modulated in time at a resonant transition frequency between states.  The interaction is described by the minimal-coupling Hamiltonian (a.u.)~\cite{Sakurai.1967}
\begin{equation}
\hat{H}=\mathbf{A}(\hat{\bf{r}}) \cdot \hat{\mathbf{p}} + \mathbf{A}(\hat{\bf{r}}) \cdot \mathbf{A}(\hat{\bf{r}}) /2, \label{eq:eq1}
\end{equation}
\noindent where $\hat{\mathbf{p}}$ is the electron's momentum operator and $\mathbf{A}$ the vector potential of the laser electric field.  In this work, the  $\mathbf{A} \cdot\mathbf{A}$ (ponderomotive) interaction drives atomic transitions, rather than the usual $\mathbf{A} \cdot \hat{\mathbf{p}}$ interaction, leading to flexible selection rules~\cite{Knuffman.2007} and expanding the range of transitions available for precision measurement.

\begin{figure}[t]
\begin{centering}
\includegraphics[width=3.3in]{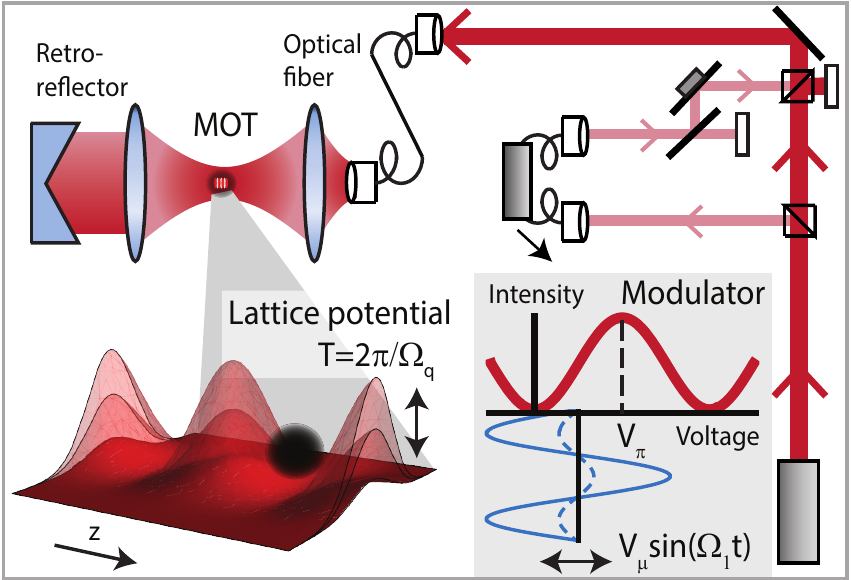}
\caption{Experimental set-up.  A Mach-Zehnder interferometer splits and re-combines two 1064-nm c.w. beams.  One beam is amplitude-modulated by a fiber-based electro-optic modulator (EOM), driven at frequency $\Omega_1$ (inset).  The optical lattice is formed by retro-reflecting and focusing the re-combined beam into a $^{85}$Rb magneto-optical trap (MOT).  The lattice potential is modulated with period $T$.} \label{fig1}
\end{centering}
\end{figure}

In Fig.~\ref{fig1} we show the experimental set-up.  See \cite{Moore.2015} for details.  A continuous-wave (c.w.) 1064-nm laser beam is split by a Mach-Zehnder interferometer into a low-power and higher-power beam.  The low-power beam is amplitude-modulated via an EOM driven by a microwave signal with voltage amplitude $V_{\mu}$ and frequency $\Omega_1$.  This beam is coherently re-combined with the unmodulated (higher-power) beam to parametrically amplify the modulation sidebands [the radical term in Eq.~\eqref{eq:eq3}].  We form a standing-wave optical lattice in the atom-field interaction region by retro-reflecting the lattice beam.   Using a lattice inversion technique~\cite{Anderson.2011}, we laser-excite Rydberg atoms such that their center-of-mass locations are either at lattice intensity minima or maxima.  Intensity modulation of the lattice then results in a time-periodic atom-field ponderomotive interaction with a leading quadratic dependence on position.

\begin{figure}[b]
\begin{centering}
\includegraphics[width=3.3in]{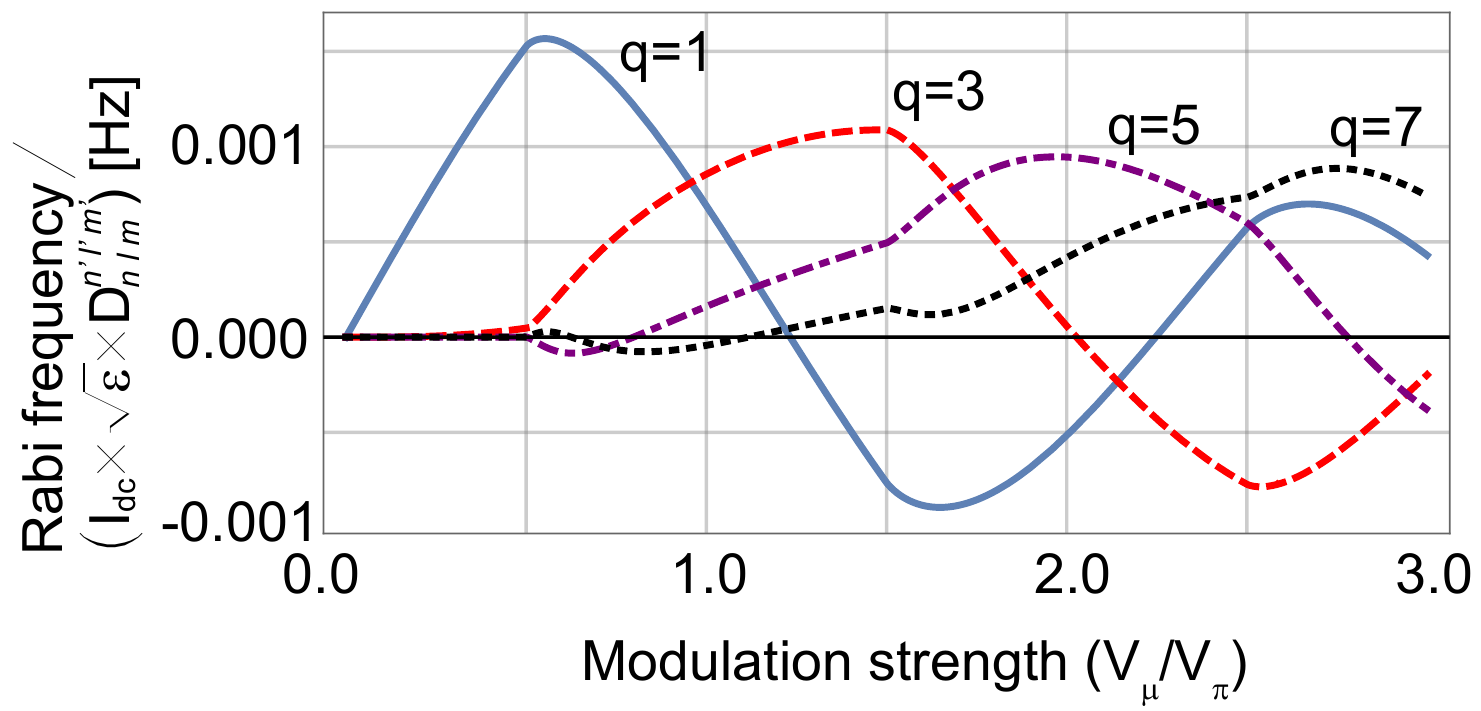}
\caption{Expected scaled Rabi frequencies of the $q$th harmonics of the lattice modulation, as a function of $V_{\mu}$ (units of $V_\pi$)~\cite{evenharmonics}.  See text for y-axis.} \label{fig2}
\end{centering}
\end{figure}

Temporal harmonics in the lattice modulation drive atomic transitions at frequencies $\Omega_q=q \Omega_1$, where $q$ is an integer.  The inset in Fig.~\ref{fig1} illustrates how this is achieved.   The EOM offset voltage is set to $V_\pi/2$, where $V_\pi$ is the voltage difference between minimum and maximum transmission.  To modulate at the fundamental frequency  $\Omega_1$, one has $ V_{\mathrm{\mu}} \lesssim V_\pi/2$ (dashed blue line).  To access higher harmonics, $ V_{\mathrm{\mu}}$ is increased (solid blue line).  The resultant frequency upconversion is described by 

\begin{align}
\frac{I_{\mathrm{inc}}}{I_{\mathrm{dc}}} &= 1 + 2 \left(\frac{\eta}{2}+\eta \sum_{q=1, 3, 5 \ldots}^{\infty} \mathrm{J}_{\mathrm{q}} \left(\frac{\pi V_{\mathrm{\mu}}}{V_\pi} \right) \sin \left(q \Omega_1 t\right)\right)^{1/2}\nonumber \\
&\qquad + \eta \left(\frac{1}{2}+\sum_{q=1, 3, 5 \ldots}^{\infty} \mathrm{J}_{\mathrm{q}} \left(\frac{\pi V_{\mathrm{\mu}}}{V_\pi} \right) \sin \left(q \Omega_1 t\right) \right), \label{eq:eq3}
\end{align}

\noindent the incident intensity $I_{\mathrm{inc}}$ at the location of the atoms, scaled by $I_{\mathrm{dc}}$ (the intensity of the unmodulated high-power beam).  Here, $\eta$ is the power ratio between the modulated and unmodulated beams in the interferometer.  A Fourier analysis of Eq.~\eqref{eq:eq3} with $\eta=0.0077$ (typical experimental value) leads to the Rabi frequency for the $q$th harmonic as a function of modulation strength $V_{\mu}/V_\pi$ (Fig.~\ref{fig2}).  The Rabi frequency is scaled by $I_{\mathrm{dc}}$ (in units of W/m$^2$); $\sqrt{\varepsilon}$, where $\varepsilon$ is the ratio of the return and incident lattice-beam intensities ($\varepsilon =0.09$ experimentally); and $D_{n, l, m}^{n', l', m}$.  The unitless, normalized matrix elements for the spatial coupling,  $D_{n, l, m}^{n', l', m}$, between Rydberg levels in the standing wave have been obtained in~\cite{Knuffman.2007}.  For our lattice intensities and transitions, the Rabi frequencies are near 100~kHz.

The Rabi frequencies exhibit a sinusoidal dependence on position in the optical lattice~\cite{Moore.2015}; in Fig.~\ref{fig2} we display the maximum Rabi frequency.  The Rabi frequency for even-parity transitions (as in our case) is maximal for atoms at lattice intensity maxima and minima, where the leading spatial dependence of the ponderomotive potential is quadratic.  Since the peak Rabi frequencies in Fig.~\ref{fig2} drop slowly as a function of $q$, unlike in typical nonlinear spectroscopy, we do not need to increase $I_{\mathrm{dc}}$ to realize higher-order transitions, thereby avoiding increased light shifts.  We utilize Fig.~\ref{fig2} to determine the $V_\mu$ needed to achieve a high Rabi frequency at the harmonic order of interest.  Atom-field interaction times are $\approx 10~\mu$s. We target $S \rightarrow S$ transitions because they are insensitive to the MOT magnetic field (which is always on).  Under traditional electric-dipole selection rules, these transitions would not be allowed in first-order perturbation theory.  However, because we employ ponderomotive spectroscopy, typical selection rules do not apply~\cite{Knuffman.2007}.

\setlength{\textfloatsep}{20pt}

\begin{figure}[b]
\begin{centering}
\includegraphics[width=3.3in]{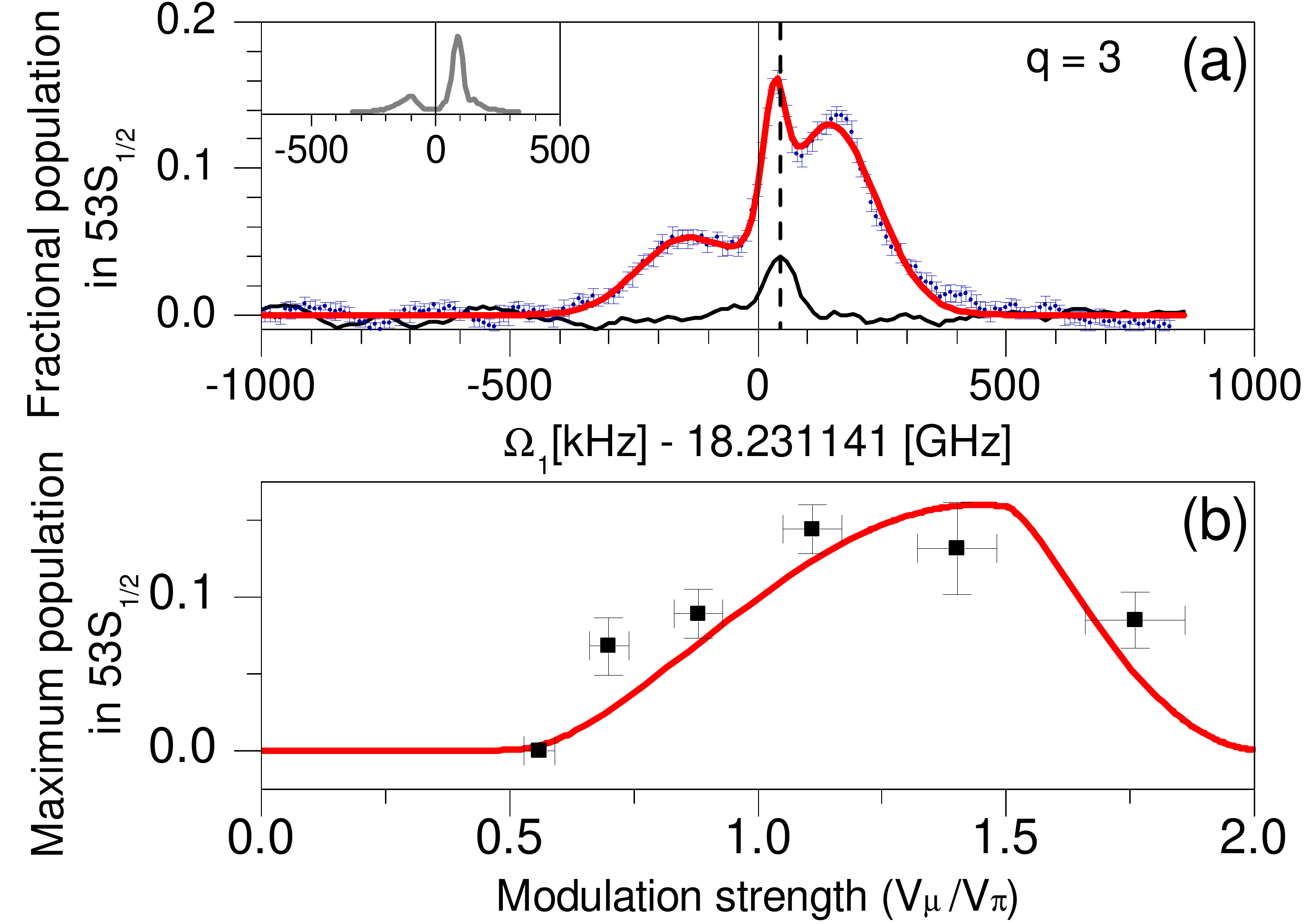}
\caption{Third harmonic drive.  (a) Population in $53S_{1/2}$ as a function of lattice modulation frequency $\Omega_1$.  Data are a smoothed average of 10 scans, 200 measurements each.  Error bars, s.e.m.  Red curve, triple-Gaussian fit.  Vertical black line, line center.  Black dashed line, center location of a two-photon microwave spectroscopy measurement (black curve).  Inset, simulation results.  (b) Peak $53S_{1/2}$ population as a function of modulation strength $V_\mu$.  Each data point represents the peak height for an average of 6-10 scans with 200 measurements each.  Vertical (horizontal) error bars, peak height ($V_\pi$) uncertainty. Red curve, proportional to the square of the $q=3$ curve in Fig.~\ref{fig2}.} \label{fig3}
\end{centering}
\end{figure}

In Fig.~\ref{fig3} we demonstrate driving the ponderomotive transition $52S_{1/2} \rightarrow 53S_{1/2}$ via the third harmonic ($q=3$) of the lattice modulation.  For this transition, $D_{52S}^{53S} = 0.19$, which is near the maximal value for $nS \rightarrow (n+1)S$ transitions, $D_{58S}^{59S} = 0.215$.  The inset in Fig.~\ref{fig3}(a) shows that most spectral features can be reproduced by a semi-classical simulation~\cite{Moore.2015}.  The simulation allows us to attribute the two outermost peaks to atoms that are anti-trapped (red-detuned peak) or trapped (blue-detuned peak) in the lattice.  The simulation does not reproduce the sharp central peak in the experimental spectrum.  This peak may be due to imperfections in the optical lattice potential, which could arise from the lattice return beam encountering multiple interfaces that degrade the quality of the return focal spot.

To determine the line center, we fit the smoothed, averaged spectral line in Fig.~\ref{fig3}(a) to a triple-Gaussian.  To eliminate most of the trap-induced systematic shifts, we take the mean of the center locations of the outermost peaks.  Assuming that the outermost peaks correspond to equal but opposite extremum light shifts (as would be the case in a perfect optical lattice), the mean provides a measurement of the line center with much-reduced light shift.  See Table~\ref{table1} for a comparison between our measured result, the calculated line center, and a two-photon microwave spectroscopy reference measurement.

In Fig.~\ref{fig3}(b) we show the dependence of the $53S_{1/2}$ excited-state population on modulation strength $V_\mu/V_\pi$.  The behavior qualitatively agrees with the square of the Rabi frequency for the $q=3$ curve in Fig.~\ref{fig2}, plotted for comparison.  In particular, significant excited-state populations only occur beyond a threshold value of $V_\mu / V_\pi \approx 0.6$.  Fig.~\ref{fig3}(b) reinforces that the spectrum in Fig.~\ref{fig3}(a) is due to the nonlinear $q=3$ component of the lattice modulation.

\begin{figure}[t]
\begin{centering}
\includegraphics[width=3.3in]{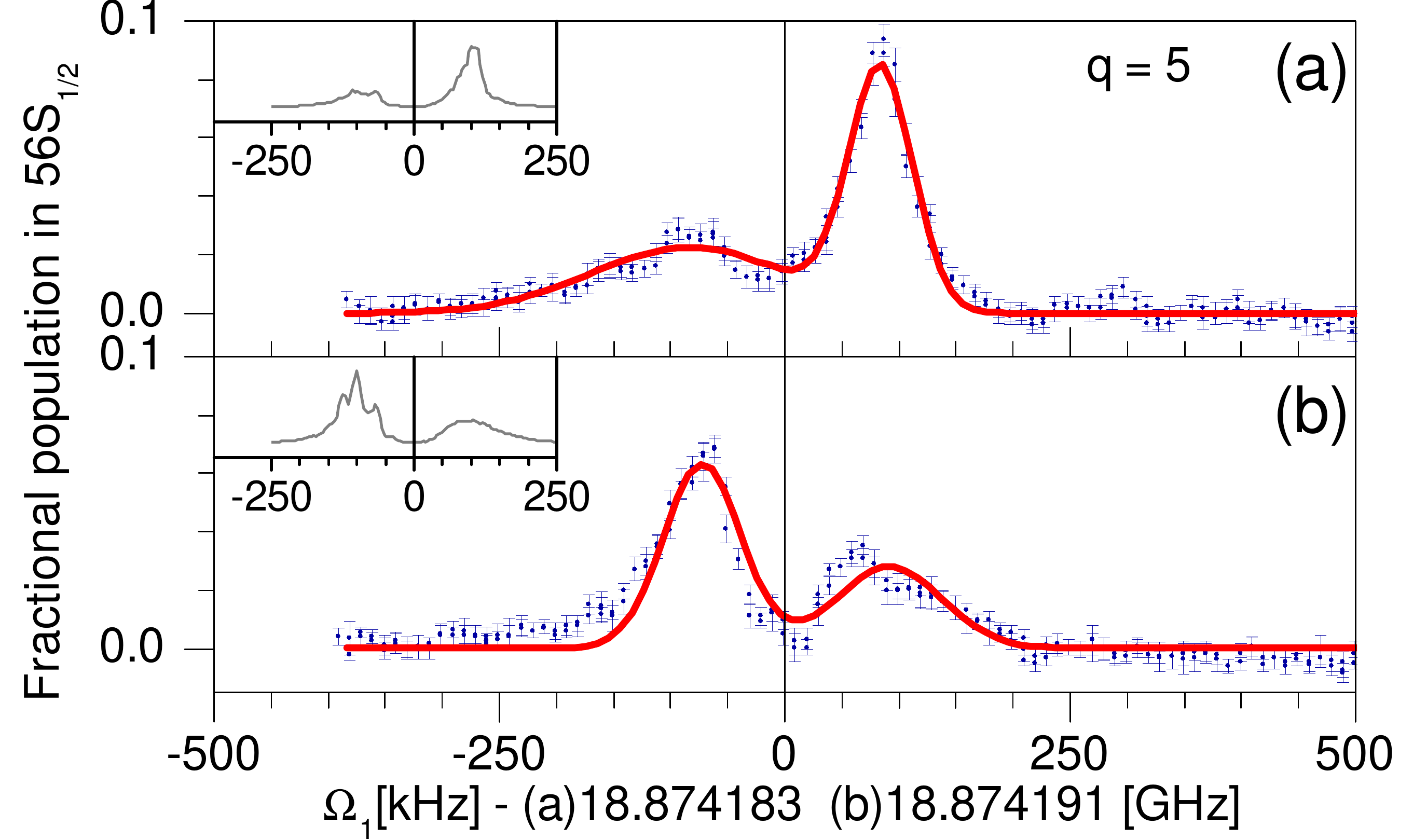}
\caption{Fifth harmonic drive. (a), (b) Population in $56S_{1/2}$ as a function of lattice modulation frequency $\Omega_1$ for Rydberg atoms prepared at lattice intensity (a) minima (b) maxima.  Data are a smoothed average of (a) 18 (b) 10 scans, 200 measurements each.  Error bars, s.e.m.   Red curve, double-Gaussian fit.  Vertical black line, line center.  Insets, simulation results.} \label{fig4}
\end{centering}
\end{figure}

To reach higher-frequency transitions, it is desirable to drive transitions via even higher harmonics.  In Fig.~\ref{fig4} we demonstrate the ponderomotive transition $54S_{1/2} \rightarrow 56S_{1/2}$, which has $D_{54S}^{56S} = 0.08$, via the fifth harmonic ($q=5$).  The spectra in Figs.~\ref{fig4}(a), (b) are reproduced accurately by the simulations (insets).  As before, the simulations indicate that the red-detuned (blue-detuned) peak represents atoms that are anti-trapped (trapped) in the lattice.  In the figure, we also demonstrate the relation between Rydberg-atom position and light-shift polarity by either preparing the Rydberg atoms near an intensity minimum [Fig.~\ref{fig4}(a)], which results in mostly trapped atoms with positive light shifts; or maximum [Fig.~\ref{fig4}(b)], which results in mostly anti-trapped atoms with negative light shifts~\cite{Anderson.2011}.

To determine the line center, we fit the spectra in Figs.~\ref{fig4}(a),(b) to a double-Gaussian.  We take the mean of the center locations of the outermost peaks, yielding a measurement in which the systematic light shifts mostly cancel.  See Table~\ref{table1} for results.  Significantly, the $q=5$ transition in Fig.~\ref{fig4} has a transition frequency of about 94.4~GHz.  Hence, we are approaching the sub-THz regime, which is important for improving $\Delta \nu/\nu$ in precision frequency measurements.

In spectroscopy, magic-wavelength lattices play an important role because they allow for probing atoms in a light trap while avoiding systematic line-shifts due to the trap~\cite{Ye.2008}.  In a ponderomotive lattice, a magic condition occurs when a Rydberg atom becomes comparable in size to the lattice period, at which point the lattice-induced shifts of certain upper and lower Rydberg levels are nearly the same~\cite{Dutta.2000} and cancel.  In a 1064-nm lattice, this phenomenon occurs for Rb $nS_{1/2}$ atoms with lower- and upper-level principal quantum numbers symmetric about $69.5$.  Furthermore, for principal quantum numbers between $n = 66 - 73$ the sign of the effective polarizability of the atoms is reversed, indicating that Rydberg atoms are attracted to lattice-field maxima.  This case differs from the more typical case where Rydberg atoms are attracted to field minima.

\begin{figure}[t]
\begin{centering}
\includegraphics[width=3.3in]{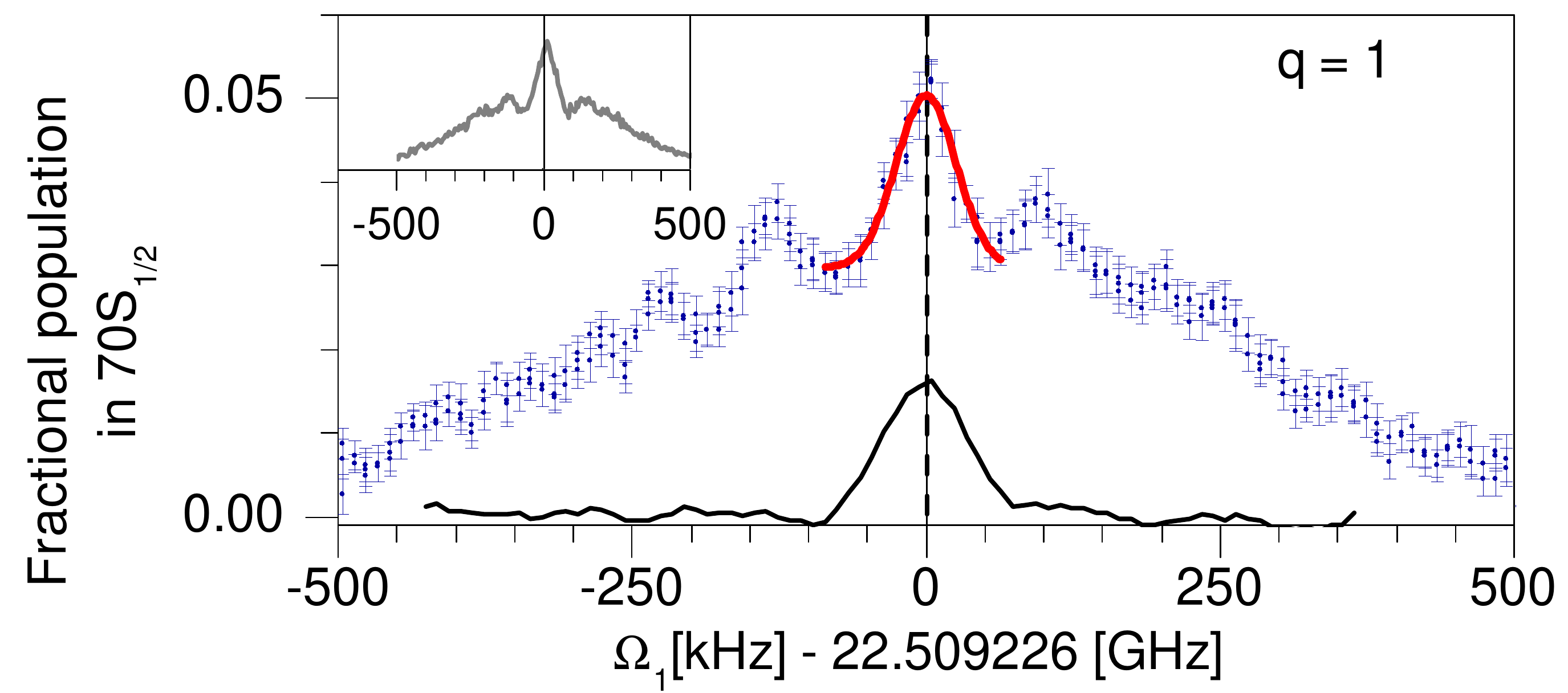}
\caption{Magic condition, fundamental.  Population in $70S_{1/2}$ as a function of lattice modulation frequency $\Omega_1$.  Data are a smoothed average of 18 scans, 200 measurements each.  Error bars, s.e.m.  Red curve, single-Gaussian fit.  Vertical black line, line center.  Black dashed line, center location of a two-photon microwave spectroscopy measurement (black curve).} \label{fig5}
\end{centering}
\end{figure}

In Fig.~\ref{fig5} we show the results of driving the magic transition $69S_{1/2} \rightarrow 70S_{1/2}$, for which $D_{69S}^{70S} = 0.13$, at the fundamental frequency ($q =1$).  The trap depths (and corresponding light shifts) for both states are about the same (2.2\% of the free-electron ponderomotive trap depth).  The magic condition results in an experimental lineshape that is symmetric and has a narrow central feature.  These characteristics are well-reproduced by the simulation (inset), which supports our interpretation of the signal.  To determine the line center, we fit the central feature, expected to have nearly-zero systematic light shift, to a single Gaussian.  As shown in Table~\ref{table1}, this measurement is determined to within 2~kHz (statistical uncertainty).

\begin{figure}[t]
\begin{centering}
\includegraphics[width=3.3in]{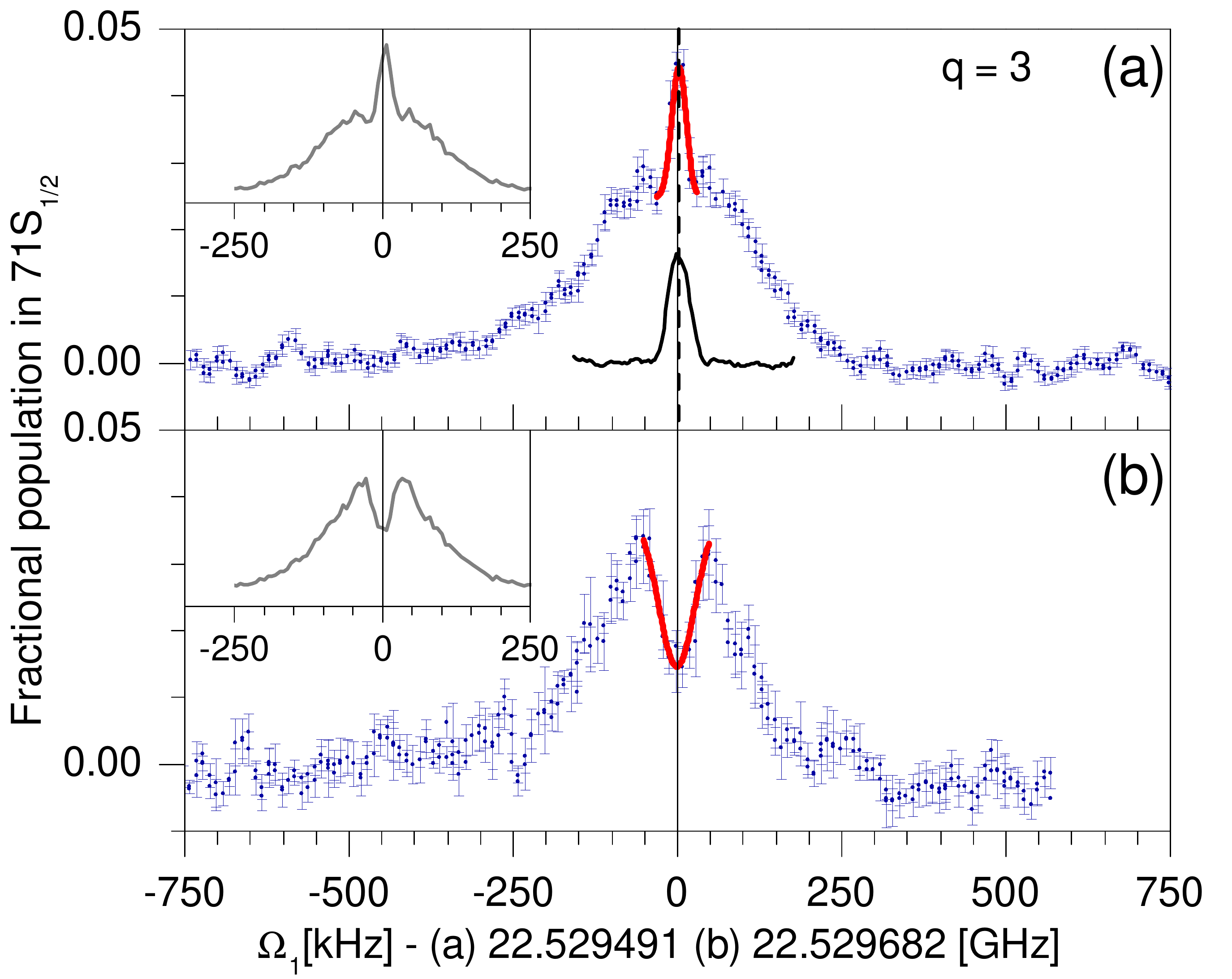}
\caption{Magic condition, third harmonic.  Population in $71S_{1/2}$ as a function of lattice modulation frequency $\Omega_1$.  Data are a smoothed average of (a) 12 (b) 5 scans, 200 measurements each.  Error bars, s.e.m.  Red curve, single-Gaussian fit.  Vertical black line, line center.  (a) Lattice not inverted.  Black dashed line, center location of a two-photon microwave spectroscopy measurement (black curve). (b) Lattice inverted.} \label{fig6}
\end{centering}
\end{figure}

For Fig.~\ref{fig6} we show a combination of both magic-condition and nonlinear ponderomotive spectroscopy, which has the greatest potential to improve $\Delta \nu / \nu$.  Here, we drive a magic ponderomotive transition $68S_{1/2} \rightarrow 71S_{1/2}$, which has $D_{68S}^{71S} = 0.08$, via the third harmonic ($q=3$).  The trap depths of the $68S_{1/2}$ and $71S_{1/2}$ states are both approximately 2.0\% of the free-electron ponderomotive trap depth.  In Fig.~\ref{fig6}(a), we prepare the Rydberg atoms at lattice intensity maxima and observe a narrow central peak on top of a broad, symmetric distribution; this central peak agrees very well with the result of a two-photon microwave spectroscopy measurement, also shown.  In Fig.~\ref{fig6}(b), we prepare the atoms at lattice intensity minima and observe a narrow central dip on top of the broad distribution.  In both cases in Fig.~\ref{fig6}, the symmetry of the lineshapes agrees with our expectation for magic lattices.  The spectral features are reproduced very well by the simulations (insets); the simulations also show that a central peak (dip) corresponds to trapped (anti-trapped) atoms.

Because both Figs.~\ref{fig5}-\ref{fig6} correspond to a magic condition, we expect the respective frequency measurements to have nearly-zero light shift.  We attribute the $\approx 200$~kHz discrepency observed between Fig.~\ref{fig6}(a)~and~(b) to differing residual electric fields, which may have caused different systematic DC shifts between the data sets (which were taken on different days).

In Fig.~\ref{fig5} (and to a lesser extent in Fig.~\ref{fig6}), we note small oscillations near the central peaks.  We attribute these to the effect of atoms running over the lattice potential wells, which are very shallow for the magic-condition lattices.  For our atom temperature, atoms typically traverse multiple lattice potential wells during the modulation interaction time.  Since the modulation-induced Rabi frequency flips sign twice per lattice period, a rotary-echo-like effect~\cite{RotEcho} occurs.  This causes periodic structures when varying the detuning.

An important finding in both Figs.~\ref{fig5}-\ref{fig6} is that the central peaks in the lattice-modulation spectra and the peaks in the two-photon reference spectra are at Fourier-limited resolution.  Therefore, precision measurements made via magic-condition ponderomotive spectroscopy and traditional microwave spectroscopy can have similar spectral resolution.  However, ponderomotive spectroscopy allows us to access a wide variety of typically-forbidden transitions at high frequencies.

\setlength{\abovecaptionskip}{8pt}
\setlength{\belowcaptionskip}{-5pt}
\setlength{\textfloatsep}{20pt}

\begin{table}
\begin{tabular}
{l l l l l}
\hline \hline
  & & $q \Omega_1$ (GHz) & $\nu_\mathrm{calc}$  (GHz) & $2 \nu_\mathrm{2p}$ (GHz) \\ \hline
  $52S_{1/2} \rightarrow 53S_{1/2}$& & \multirow{2}{*}{54.693423(15)} &  \multirow{2}{*}{54.693577(5)} &  \multirow{2}{*}{54.693556(6)} \\
  \hspace*{1em} $q=3$&&&& \\
$54S_{1/2} \rightarrow 56S_{1/2}$ & a. & 94.370915(15) & 94.371060(8) & \\
  \hspace*{1em} $q=5$ &  b. & 94.370955(15) & 94.371060(8) & \\
  $69S_{1/2} \rightarrow 70S_{1/2}$ & &  \multirow{2}{*}{22.509226(2)} &  \multirow{2}{*}{22.509227(1)} &  \multirow{2}{*}{22.509226(2)} \\
   \hspace*{1em} $q=1$, magic &&&& \\
 $68S_{1/2} \rightarrow 71S_{1/2}$ & a. & 67.588473(2) & 67.589048(4) & 67.588478(1) \\
  \hspace*{1em} $q=3$, magic &  b. & 67.589046(9) & 67.589048(4) & \\
\hline \hline
\end{tabular}
\caption{Summary of results.  All measurement uncertainties, statistical.  See text for details.} \label{table1}
\end{table}

Frequency-measurement results are summarized in Table~\ref{table1}, where we list transition frequencies ($q \Omega_1$) measured in the modulated lattice, expected transition frequencies ($\nu_\mathrm{calc}$) calculated using quantum defects from \cite{EITquantumdef}, and reference measurements ($2\nu_\mathrm{2p}$) obtained via two-photon microwave spectroscopy in the absence of a lattice.  Overall, agreement is quite satisfactory.  Statistically significant shifts in the measured results from the calculated values are negative, indicating possible Stark shifts due to residual DC electric fields.  In the magic-condition cases, the statistical uncertainty in $q \Omega_1$ is of the same order of magnitude as the uncertainty in $2\nu_\mathrm{2p}$.  With improvement of stray electric field control, the modulated lattice should be suitable to measure atomic transition frequencies that are difficult or forbidden in microwave spectroscopy.

In conclusion, we have performed ponderomotive spectroscopy of Rydberg atoms by employing higher harmonics of the lattice modulation as well as magic conditions of the lattice.  We have reached transitions near the sub-THz regime, with the potential to exceed that limit, using lower-frequency microwave sources.  Using magic lattices, we have demonstrated Fourier-limited spectral lines with a small $\Delta \nu$, as well as reduced light shifts.  Being able to access sub-THz atomic transitions with low $\Delta \nu$ and free from light shifts will improve the fractional resolution ($\Delta \nu/\nu$) of transition frequency measurements.  Applications include precision measurement of atomic characteristics~\cite{Han.2006} and physical constants (e.g. the Rydberg constant~\cite{Kleppner.1997}, which could lead to verification of the proton size~\cite{Pohl.2010}).

This work was supported by NSF Grant No. PHY-1205559, NIST Grant No. 60NANB12D268, and NASA Grant No. NNN12AA01C.
*kaimoore@umich.edu

\bibliographystyle{apsrev4-1}
\bibliography{References}

\begin{thebibliography}{14}
\bibitem [{eve()}]{evenharmonics}%
  \BibitemOpen
  \href@noop {} {}\bibinfo {note} {We do not plot even-order harmonics in
  Fig.~2 because in our set-up the modulated EOM voltage is centered at
  $V_{\pi}/2$ (see Fig.~1), and therefore odd orders dominate the modulation
  spectrum.}\BibitemShut {Stop}%
\end{thebibliography}

\end{document}